\documentclass[runningheads]{llncs}
\usepackage[T1]{fontenc}
\usepackage{graphicx}
\usepackage{todonotes} 
\usepackage{adjustbox}
\usepackage{multirow}
\usepackage{booktabs}
\usepackage{hyperref}
\usepackage{amsmath}
\usepackage{amsfonts}
\usepackage{amssymb}
\usepackage{pifont}
\usepackage{xcolor, colortbl}

\definecolor{cblue}{rgb}{0,0.5,0.8}
\definecolor{cgreen}{rgb}{0,0.7,0.6}
\definecolor{cred}{rgb}{0.968,0.145,0.521}
\newcommand{\xmark}{\ding{55}}%

\newcommand{\rtr}[1]{{\scriptsize\color{cred}$\blacktriangledown$ #1}}

\newcommand{\gtr}[1]{{\scriptsize\color{cred}$\blacktriangle$ #1}}
\newcommand{\notr}[1]{{\scriptsize\color{cgreen}$\blacktriangleright$ #1}}
\newcommand{\btr}[1]{{\scriptsize\color{cblue}$\uparrow$ #1}}

\definecolor{turq}{rgb}{0.286, 0.658, 0.678}
\definecolor{kingsred}{rgb}{0.882, 0.136, 0.097}

\begin{document}
\title{Bias in Unsupervised Anomaly Detection in Brain MRI}
%
%
\author{Cosmin I. Bercea \inst{1,2,5} \and
Esther Puyol-Antón \inst{5} \and 
Benedikt Wiestler \inst{3} \and
Daniel Rueckert \inst{1,3,4} \and
Julia A. Schnabel \inst{1,2,5} \and
Andrew P. King \inst{5}
}

\authorrunning{C.I. Bercea et al.}
\institute{Technical University of Munich, Germany \and
Helmholtz AI and Helmholtz Center Munich, Germany \and 
Klinikum Rechts der Isar, Munich, Germany \and 
Imperial College London, London, UK  \and 
King’s College London, London, UK 
}
%
\maketitle              
\begin{abstract}
Unsupervised anomaly detection methods offer a promising and flexible alternative to supervised approaches, holding the potential to revolutionize medical scan analysis and enhance diagnostic performance. 

In the current landscape, it is commonly assumed that differences between a test case and the training distribution are attributed solely to pathological conditions, implying that any disparity indicates an anomaly. However, the presence of other potential sources of distributional shift, including scanner, age, sex, or race, is frequently overlooked. These shifts can significantly impact the accuracy of the anomaly detection task. Prominent instances of such failures have sparked concerns regarding the bias, credibility, and fairness of  anomaly detection.

This work presents a novel analysis of biases in unsupervised anomaly detection. By examining potential non-pathological distributional shifts between the training and testing distributions, we shed light on the extent of these biases and their influence on anomaly detection results. Moreover, this study examines the algorithmic limitations that arise due to biases, providing valuable insights into the challenges encountered by anomaly detection algorithms in accurately learning and capturing the entire range of variability present in the normative distribution. Through this analysis, we aim to enhance the understanding of these biases and pave the way for future improvements in the field. Here, we specifically investigate Alzheimer's disease detection from brain MR imaging as a case study, revealing significant biases related to sex, race, and scanner variations that substantially impact the results. These findings 
\end{abstract}
%
%
%
%
\section{Introduction}
Unsupervised anomaly detection (UAD) methods have gained significant attention in the medical image analysis research literature due to their potential to identify anomalies without the need for labeled training data. However, recent literature has shown that UAD methods are vulnerable to non-pathological out-of-distribution (OoD) data~\cite{heer2021the}. As a result, notable failures in such approaches have raised concerns regarding bias and fairness in their evaluation. For example, Meissen et al.~\cite{meissen2022domain} presented cases where polyp detection algorithms achieved excellent performance even when the actual polyps were removed from the error maps. Similarly, Bercea et al.~\cite{bercea2022we} demonstrated nearly perfect OoD detection using popular reconstruction-based methods that solely relied on analyzing background pixels. Moreover, a predominant focus in the recent literature on UAD has been on the detection of hyper-intense lesions in brain MRIs~\cite{chen2020unsupervised,kascenas2022denoising,pinaya2021unsupervised,zimmerer2019unsupervised}. A recent study demonstrated that many reconstruction-based methods struggled to generalize to other types of anomalies, indicating a bias in the algorithmic performance~\cite{bercea2022ra}. In light of these notable failures, it is essential to thoroughly investigate these concerns to enable the development of more robust and reliable models that exhibit fair and unbiased behavior.

There has been relatively little research into bias in UAD techniques. This is in contrast to other medical imaging applications, where in recent years there has been an increasing focus on bias and fairness. For example, Gichoya et al.~\cite{Gichoya2022} demonstrated the presence of race-based distributional shifts across several imaging modalities, highlighting the potential for bias when training models with imbalanced data. Additionally, studies such as Larrazabal et al.~\cite{Larrazabal2020} and Seyyed et al.~\cite{Seyyed2021} have examined bias in chest X-ray classification, Guo et al.~\cite{Guo2021} reviewed work on biases in skin cancer detection algorithms, and Puyol-Anton et al.~\cite{PuyolAnton2022} have identified race bias in cardiac MR segmentation. In recent years, there have been several studies that have examined biases in neuroimaging data~\cite{Stanley2022a,Stanley2022,Wang2023},including the task of Alzheimer's disease (AD) detection~\cite{Petersen2022}.

However, these studies have all focused on supervised approaches. In contrast, our study specifically investigates unsupervised models, which are underpinned by the need to learn the normative training distribution and thus could be more susceptible to biases. These studies underscore the importance of addressing bias in medical applications and sand pave the way for further exploration in UAD. 
The motivation for studying bias in UAD is twofold. First, given that most experimental setups in the research literature have involved distinct data sources for healthy and pathological distributions, it is essential to analyze potential shifts to ensure fair evaluations and prevent correlations that can significantly skew the performance of these methods. Second, as UAD models strive to represent the entire variability of the normative distribution and effectively identify anomalies by isolating pathological shifts, it becomes increasingly more important to identify their algorithmic limitations.

In this work, we investigate both types of biases in anomaly detection, aiming to fill the gap in this important research area. Our focus is on a case study of AD detection from brain MR images. In summary, our main contributions are: 
\begin{itemize}
    \item[$\bullet$] To the best of our knowledge, this work represents the first comprehensive investigation into the biases present in UAD.
    \item[$\bullet$] Through rigorous analysis, we have uncovered evidence of scanner, sex, race, and metrics biases that significantly impact the performance of UAD.
    \item[$\bullet$] We examined other factors like age and brain volume but found no additional correlations for the observed performance drops.
\end{itemize}
 By shedding light on these biases, we strive to enhance the reliability, fairness, and effectiveness of anomaly detection methods in medical imaging, ultimately benefiting both healthcare providers and patients.
\section{Materials and Methods\label{sec::data}}
\begin{table}[t!]
    \centering
    \setlength{\tabcolsep}{8pt}
    \caption{\textbf{Datasets.}
We present the data splits utilized in our experiments. The abbreviations (Abv.) are linked to the experiments in~\autoref{tab::baseline}. We refer to the healthy training distribution as "Control", the healthy cohort from a different distribution as "Healthy". "Alzheimer's (AD)" repsents the pathology set. We mark in \colorbox{cblue!20}{blue} the shifts in distribution compared to the control set. \label{tab::data}}
    \begin{adjustbox}{width=0.95\linewidth,center} 
        \begin{tabular}{l|l | c c c c c}
            \toprule	    
            Abv. & Dataset & \#Scans & Group & Ethnicity & Sex & Scanner \\
            \midrule
            T & Training (Control) & 434 & Control & White & Female & Siemens \\
            V& Validation (Control) & 54 & Control & White & Female & Siemens \\ \midrule \midrule
            C& Test (Control) & 122 & Control & White & Female & Siemens \\
            AD& Test (Baseline) & 131 & AD & White & Female & Siemens \\\midrule
            H1& Test (Healthy, Scanner Shift 1) & 171 & Healthy & White & Female & {\cellcolor{cblue!20}Philips} \\
            AD1& Test (Scanner Shift 1) & 73 & AD & White & Female & {\cellcolor{cblue!20}Philips} \\\midrule
            H2& Test (Healthy, Scanner Shift 2) & 70 & Healthy & White & Female & {\cellcolor{cblue!20}GE} \\
            AD2& Test (Scanner Shift 2) & 36 & AD & White & Female & {\cellcolor{cblue!20}GE} \\\midrule
            H3& Test (Healthy, Sex Shift) & 480 & Healthy & White & {\cellcolor{cblue!20}Male} & Siemens \\
            AD3& Test (Sex Shift) & 188 & AD & White & {\cellcolor{cblue!20}Male} & Siemens \\\midrule
            H4& Test (Healthy, Race Shift 1) & 103 & Healthy & {\cellcolor{cblue!20}Black} & Female & Siemens \\
            AD4& Test (Race Shift 1) & 16 & AD & {\cellcolor{cblue!20}Black} & Female & Siemens \\\midrule
            H5& Test (Healthy, Race Shift 2)& 18 & Healthy & {\cellcolor{cblue!20}Asian} & Female & Siemens \\
            AD5& Test (Race Shift 2) & 8 & AD & {\cellcolor{cblue!20}Asian} & Female & Siemens \\\midrule
     	    \bottomrule
        \end{tabular}
    \end{adjustbox}
\end{table}

\textbf{Dataset. } Data used in this study were obtained from the Alzheimer’s Disease Neuroimaging Initiative (ADNI) database~\footnote{\url{https://adni.loni.usc.edu}}. ADNI offers a rich collection of magnetic resonance imaging (MRI) scans, accompanied by comprehensive metadata including MR scanner information, and demographic factors such as age, sex, and race. This dataset offers an ideal opportunity to isolate specific factors and evaluate their impact on anomaly detection. In~\autoref{tab::data}, we present an overview of the data utilized in this paper, specifically detailing the partitioning of the ADNI dataset into different training, validation and test subsets. As can be seen, our training dataset was acquired from white females using Siemens scanners with a field strength of 3T. This choice was made to maximize the availability of training images and facilitates a more comprehensive assessment of the model's performance.\\
\textbf{UAD Method. } We utilized a state-of-the-art variational auto-encoder architecture as our UAD method\footnote{\url{https://github.com/Project-MONAI/GenerativeModels}}. This recent model incorporates advanced techniques, including perceptual and adversarial loss functions, to enhance the accuracy of image reconstructions, while constraining the latent distribution using the Kullback-Leibler divergence. \\
\textbf{Metrics. } We use a range of metrics to evaluate the performance of our method from different perspectives. To assess the reconstruction quality of the methods, we use the mean absolute error (MAE). 
To assess the anomaly detection ability of our method, we use area under the receiver operator curve (AUROC) and area under the precision-recall curve (AUPRC). Additionally, we include the subjective assessment of a clinician expert to evaluate the quality of reconstructions and the localization of anomalies. Finally, for assessing statistical significance, we used Pearson's correlation to identify the impact of potential confounders on the residual errors, and the Kolomogorov-Smirnov test to identify distributional shifts between the AD sets of the training distribution and AD sets of the target distributions. We considered results with p-values lower than 0.05 significant.

\section{Experiments and Results}
In~\autoref{sec::baseline}, we conduct a comprehensive evaluation of the proposed method under ideal conditions, where no distributional shifts other than the pathological one are expected. This evaluation provides insight into the ability of the model to accurately detect AD. Subsequently, in~\autoref{sec::bias}, we systematically introduce changes in the data distribution by modifying a single factor other than pathology, such as MRI scanner manufacturer, sex, or race and evaluate the impact of biases on the performance. Finally, in~\autoref{sec::source_bias} we perform further analysis to uncover potential causes of the performance drops. 
\begin{figure}
    \centering
    \includegraphics[width=\columnwidth]{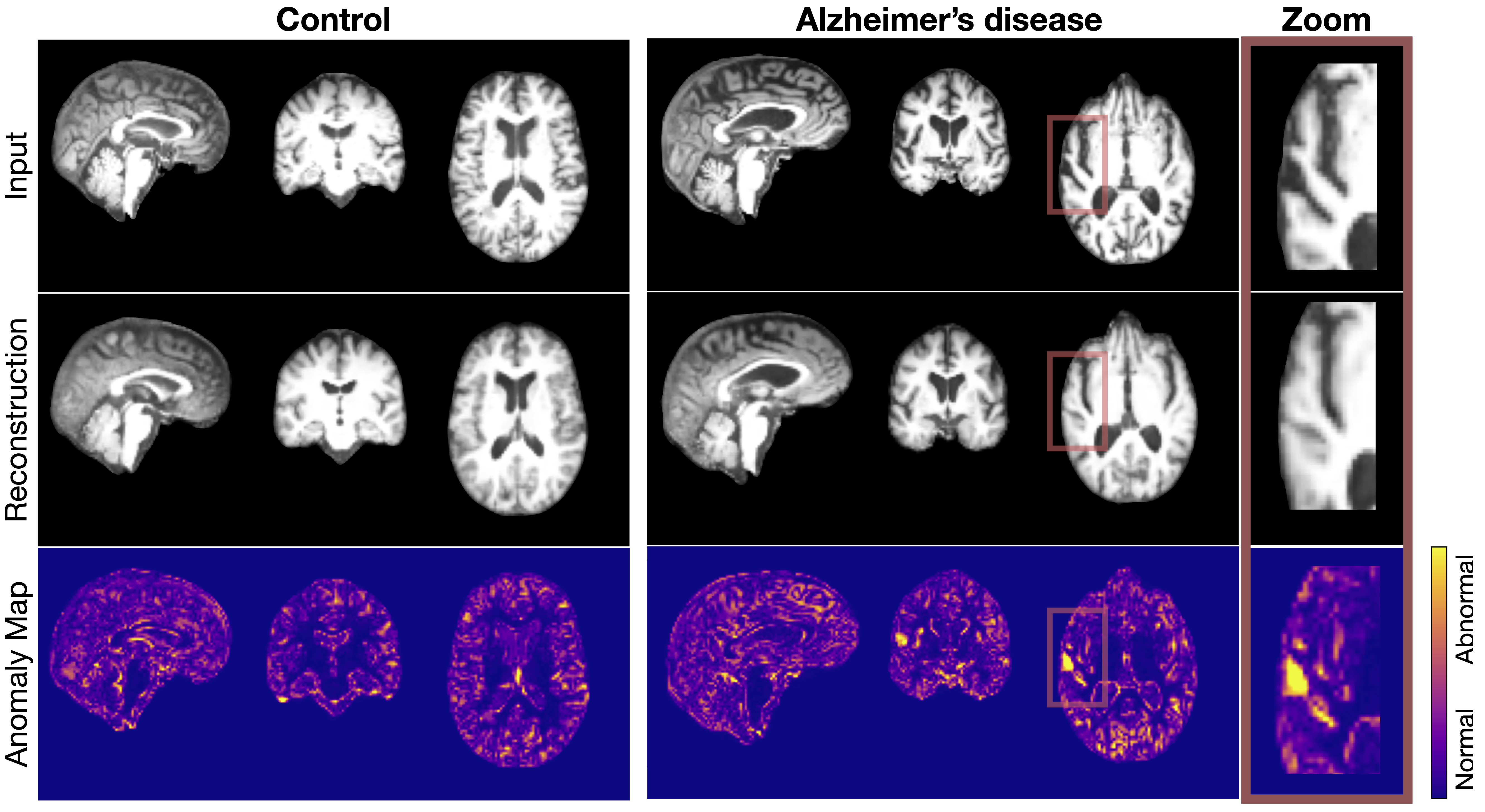}
    \caption{Qualitative results of the baseline experiment. In the case of AD (right), the atrophy in the Sylvian fissure, which is a typical feature of the disease, is reduced in the reconstruction. This leads to a clear highlight in the anomaly map.}
    \label{fig::quali}
\end{figure}
\begin{table}[t!]
    \centering
    \setlength{\tabcolsep}{6pt}
    \caption{\textbf{Bias in UAD.} The conventional approach to evaluating UAD methods involves using the control set from the training distribution (denoted as 'C' in~\autoref{tab::data}) as the healthy subjects during testing. We present these results as the Naive AD Detection. An increase~\gtr{x\%} and decrease~\rtr{x\%} in performance compared to the baseline signifies the presence of bias, while~\notr{x\%} suggests no bias. To focus solely on the methodological bias, we also report the True AD Detection, which involves using both healthy and pathological subjects from the same source at test time, such as H1/AD1.~\rtr{x\%} indicates a decrease in performance compared to the baseline caused by the distributional shift, while~\btr{x\%} demonstrates improved performance. We show the distributions of the residual errors and the visualize the bias shifts in~\autoref{fig::violin}. 
    \label{tab::baseline}}
    \begin{adjustbox}{width=\linewidth,center} 
        \begin{tabular}{l | c c c || c c c }
            \toprule	    
             & \multicolumn{3}{c||}{Naive AD Detection} & \multicolumn{3}{c}{True AD Detection}\\
            Test set & \multicolumn{3}{c||}{Evaluation \& Methodological Bias} & \multicolumn{3}{c}{Methodological Bias}\\
             & Data & AUROC $\uparrow$ &  AUPRC $\uparrow$& Data & AUROC $\uparrow$ &  AUPRC $\uparrow$
            \\\midrule
            Baseline & C/AD & 64.60 & 64.81 & C/AD & 64.60 & 64.82 \\\midrule
            Scanner (Philips) & C/AD1& 50.30~\rtr{28\%} & 41.71~\rtr{55\%} & H1/AD1 & 54.72~\rtr{18\%} & 35.23~\rtr{84\%} \\
            Scanner (GE)& C/AD2 &  60.22~\rtr{7\%} & 30.73~\rtr{111\%}  &H2/AD2 & 64.64~\notr{0\%} & 59.80~\rtr{8\%}\\
            Sex (Male) & C/AD3 &  86.68~\gtr{25\%} & 89.77~\gtr{28\%} & H3/AD3 &61.69~\rtr{5\%} & 38.39~\rtr{69\%} \\
            Race (Black)& C/AD4 & 55.53~\rtr{16\%} & 12.77~\rtr{408\%}  & H4/AD4 &56.43~\rtr{14\%} & 15.07~\rtr{330\%}\\
            Race (Asian)& C/AD5 & 65.06~\notr{1\%} & 9.65~\rtr{572\%} & H5/AD5 &73.61~\btr{12\%} & 67.59~\btr{4\%} \\
     	    \bottomrule
        \end{tabular}
    \end{adjustbox}
\end{table}
\subsection{Baseline Performance \label{sec::baseline}} 
\begin{figure}
    \centering
    \includegraphics[width=\columnwidth]{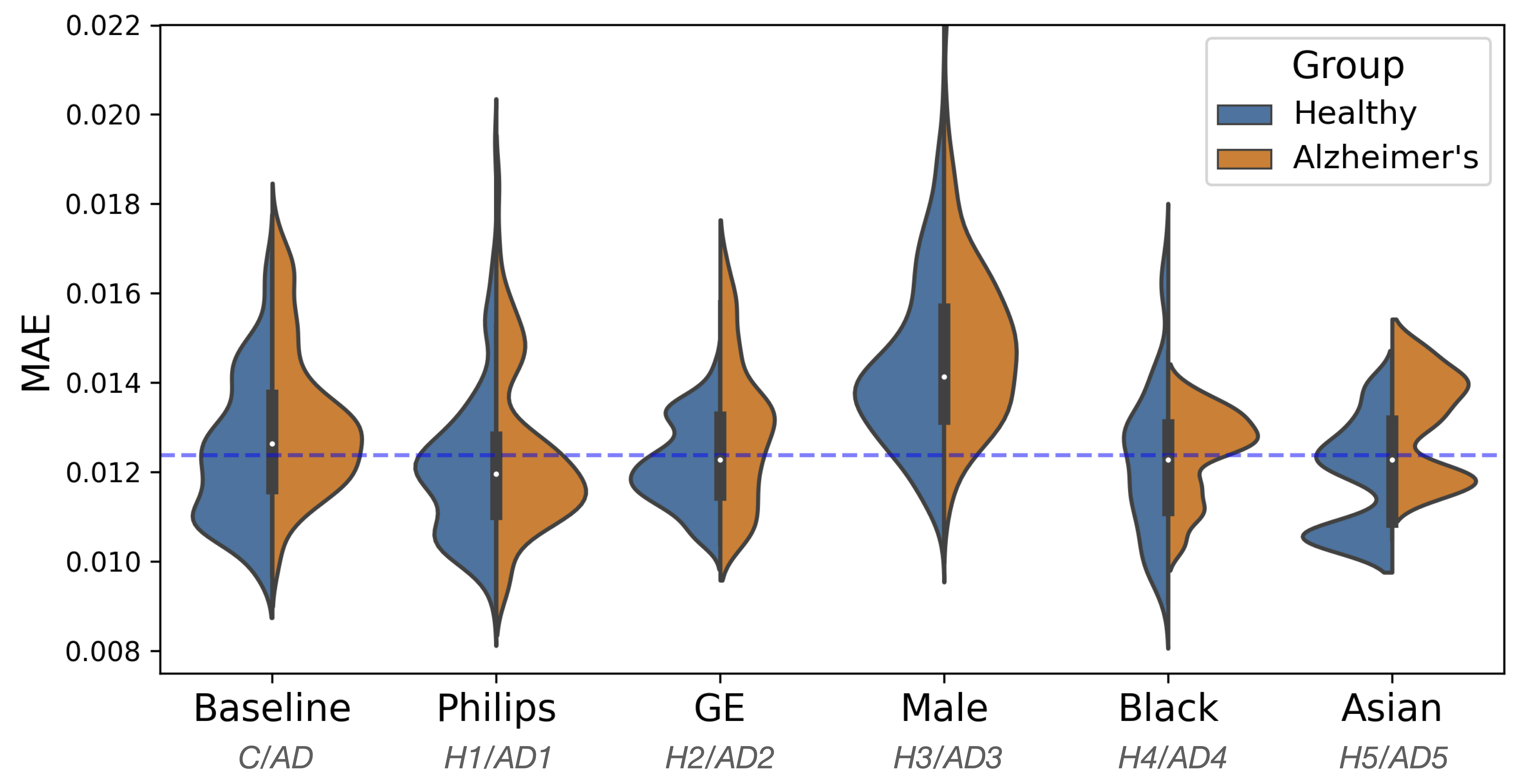}
    \caption{The distributions of residual errors for different distributional shifts demonstrates the impact of biases. Evaluation bias is characterized by a shift in the overall mean of the residual error, either above or below the mean of the training distribution (shown as a dotted line). Methodological biases are observed when comparing the distributions of healthy and AD groups within a specific shift (violin), revealing a lack of clear distinction between the two distributions. See~\autoref{tab::baseline} for numerical results.}
    \label{fig::violin}
\end{figure}

We first evaluated the baseline performance under ideal conditions, where the only factor of change was the presence of AD pathology. See~\autoref{tab::baseline} for quantitative results and~\autoref{fig::quali} for a visual example. The VAE provides detailed reconstructions, while reversing AD-related pathological atrophy, such as the dilation of the ventricles or the Sylvian fissure. Consequently, such areas are highlighted in the residual anomaly map and thus can be readily interpreted for their plausibility. The distributions plot in~\autoref{fig::violin} (Baseline) shows increased reconstruction errors for AD compared to the control set. Therefore, the method achieved moderate discriminative performance in detecting Alzheimer's pathology with AUROC and AUPRC scores of approximately 65\%. 

\subsection{Impact of Bias \label{sec::bias}}
In this section, we conducted controlled experiments to systematically investigate the impacts of distributional shifts caused by various variables, such as scanner type, sex, and race, aiming to identify potential sources of bias that might impact the performance of UAD. We summarize the results quantitatively in~\autoref{tab::baseline}, and visualize the distributions of the residual errors in~\autoref{fig::violin}.\\
We considered two distinct scenarios in our analysis. The first scenario, which we call the "naïve" approach, is commonly used in the research literature. Since there is a lack of healthy data within many publicly available datasets, most UAD use alternative sources of healthy data as controls and evaluate their performance on pathology data from a different distribution. However, this approach introduces additional distributional shifts into the evaluation process beyond the presence of pathology itself. Consequently, the evaluation is not clinically realistic and many cases of failure reported in the literature can be attributed to these confounders. In the second scenario, we aimed to address these confounders and isolate the impact of pathology by using an evaluation set that includes both healthy and pathological subjects from the same source. In doing so, we sought to assess any methodological shortcomings and evaluate the effects of bias on UAD. In both scenarios, we observed indications of significant bias stemming from factors such as scanner type, sex and race within the target groups.\\
\textbf{Domain Shifts.} First, we observed a significant domain shift in the distribution of residual errors, as depicted in~\autoref{fig::violin}. This shift manifests as a uniform vertical shift of both healthy and AD distributions compared to the baseline distribution. Note that this would lead to artificially high UAD performance when controls from the baseline distribution are employed. It is essential to recognize and address these evaluation biases, as they have the potential to strongly influence and distort the UAD results, as demonstrated in~\autoref{tab::baseline}. \\
\textbf{Metrics Bias.} Next, we examined the bias introduced by the choice of evaluation parameters. Specifically, we found that the AUROC metric tended to be too optimistic, especially when the healthy and pathological samples were highly imbalanced. In some cases the AUROC failed to recognize the presence of bias, e.g., GE scanner shifts (0\% performance difference) and Asian race shifts (1\% performance difference) in~\autoref{tab::baseline}. Instead, AUPRC emerged as a more robust measure for assessing performance, demonstrating that it is more suitable for evaluating imbalanced datasets.\\
\textbf{Scanner Shifts.} When using the conventional approach of using the training distribution control set as a reference (Naive AD Detection), we observed a performance decrease of 55\% in AURPC for the Philips scanner (C/AD1) compared to the baseline. Similarly, the GE scanner (C/AD2) showed a substantial 111\% decrease in AUPRC. To isolate methodological bias, we used both healthy and pathological distributions from the same source (True AD Detection). Interestingly, the performance on Philips (H1/AD1) still showed a considerable drop of 84\%, while for GE (H2/AD2) we only observed a minor 8\% performance drop.\\
\textbf{Sex Shifts.} 
A notable distinction between Naive and True AD detection performance occurs when examining the presence of sex bias. In the naive approach, AD detection performance for the male group increased significantly to an AUPRC of 89.77. However, a closer examination of the distribution plot shown in~\autoref{fig::violin} reveals that both the healthy and pathology distributions show higher residual errors. This finding suggests a pronounced evaluation bias associated with the naive approach. In contrast, when considering the True AD scenario, which takes into account both healthy and pathological distributions from the same source, the performance dropped dramatically to only 38.39\%. \\
\textbf{Race Shifts.} The race shift analysis revealed notable biases in anomaly detection performance. Specifically, we observed a considerable decrease in performance of 330\% when evaluating samples from the black race. Conversely, there was a slight improvement in the True AD Detection performance for subjects from the Asian race, albeit with a limited number of samples. 
\begin{table}[t!]
    \centering
    \setlength{\tabcolsep}{2pt}
    \caption{\textbf{Sources of Bias.} Statistical correlations among age, ventricular volume (VV), Hippocampal volume (HV), and whole brain volume (WBV) are examined to explore potential underlying causes for performance drops in the presence of domain shifts. Significant correlations between the analyzed confounds and increased/decreased residual errors in AD samples are denoted by a checkmark ($\checkmark$), while no correlation is indicated by a cross (\xmark). Significant shifts in the AD distributions are highlighted in $\mathbf{bold}$ and the combination of both ($\checkmark$ and $\mathbf{bold}$) is shown in \colorbox{red!10}{red}. \label{tab::corr}}
    \begin{adjustbox}{width=\linewidth,center} 
        \begin{tabular}{l  | c | c  | c | c | c | c | c | c | c | c }
            \toprule	    
             Shift $\rightarrow$ & \multicolumn{2}{c|}{Philips} & \multicolumn{2}{c|}{GE} & \multicolumn{2}{c|}{Male}& \multicolumn{2}{c|}{Black}& \multicolumn{2}{c}{Asian} \\ 
             \midrule 
            Age & \xmark &  $\mathbf{(0.224, 0.015)}$ & \xmark & $\mathbf{(0.255, 0.041)}$ & \cellcolor{red!10}{$\checkmark$} & \cellcolor{red!10}{$\mathbf{(0.346, 0.000)}$} &\xmark & (0.321, 0.082) &$\checkmark$  & (0.381, 0.172)\\
            VV  & $\checkmark$ &(0.216, 0.089) &$\checkmark$ & (0.265, 0.073) & \cellcolor{red!10}{$\checkmark$} & \cellcolor{red!10}{$\mathbf{(0.373, 0.000)}$} & \cellcolor{red!10}{$\checkmark$} & \cellcolor{red!10}{$\mathbf{(0.643, 0.002)}$} & \xmark & (0.429, 0.105)\\
            HV  & $\checkmark$ &(0.119, 0.728) & \xmark & $\mathbf{(0.403, 0.002)}$ &$\checkmark$ & (0.195, 0.063) & \cellcolor{red!10}{$\checkmark$} & \cellcolor{red!10}{$\mathbf{(0.551, 0.015)}$} &\xmark & (0.277, 0.543)\\
            WBV & $\checkmark$ &(0.230, 0.065) &$\checkmark$ & (0.207, 0.242 ) & \cellcolor{red!10}{$\checkmark$} & \cellcolor{red!10}{$\mathbf{(0.494, 0.000)}$} & \cellcolor{red!10}{$\checkmark$} & \cellcolor{red!10}{$\mathbf{(0.581, 0.008)}$} & $\checkmark$ &(0.476, 0.051)\\
     	    \bottomrule
        \end{tabular}
    \end{adjustbox}
\end{table}
\begin{figure}
    \centering
    \includegraphics[width=\textwidth]{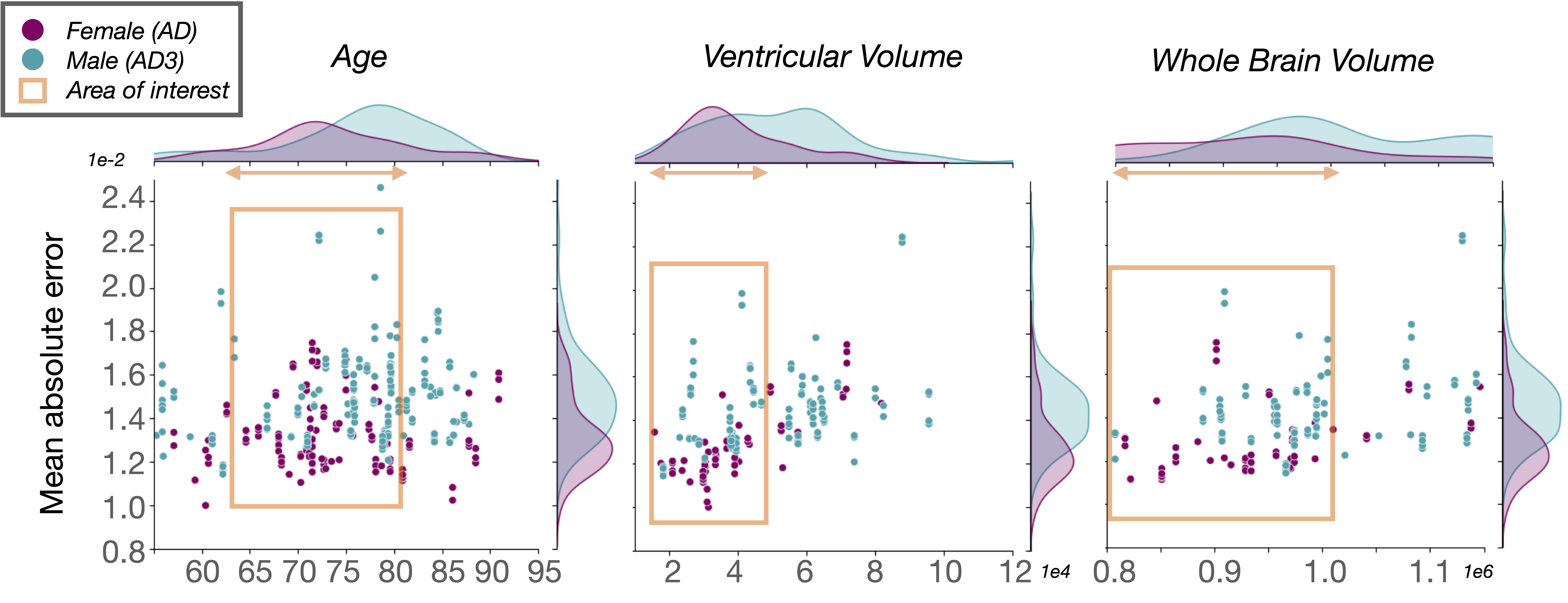}
    \caption{Visual analysis of the correlations for the sex shift reveals distribution differences in various factors compared to the AD training distribution (distribution plots on top). However, these factors do not seem to be the cause of the performance drop. A closer examination of the target area (highlighted by an orange rectangle) indicates that males have larger overall residual errors than females, suggesting an unaccounted underlying cause for the performance drops.\label{fig::corr_male}}
\end{figure}
\subsection{Sources of Bias\label{sec::source_bias}}
In this section, our objective is to explore potential causes for the observed performance drops under different shifts.~\autoref{tab::corr} presents the results of our statistical analysis, focusing on distributional shifts resulting from potential confounders, including age, ventricular volume (VV), Hippocampal volume (HV), and whole brain volume (WBV). We mark in \colorbox{red!10}{red} identified significant correlations, where a confounder significantly (according to Pearson's correlation p-values) impacts the residual errors (marked with a checkmark) and there is a significant (according to a Kolmogorov-Smirnov test) distributional shift between the training and target AD distributions (indicated in bold). To summarise, we identified significant correlations for the male and black female distributions. We further inspect the sex shift visually in~\autoref{fig::corr_male}. The analysis demonstrates elevated residual errors for males, even within the shared ranges of the analyzed confounding factors between males and females. This suggests the presence of another underlying cause beyond the factors evaluated. Further investigations, exploring additional potential confounders are necessary to uncover potential explanations and causal factors contributing to the observed performance drops in the presence of bias.
\section{Conclusion\label{sec::conclusion}}
In conclusion, our study highlights the presence of bias, including bias due to scanner, sex and race, in the performance of UAD algorithms. The results indicate that non-pathological distributional shifts can introduce significant distortions in UAD performance. These biases not only impact the overall error distribution, i.e., evaluation bias, but also affect the ability of the methods to accurately detect AD disease. It is essential to understand and address these biases in order to develop robust and reliable UAD algorithms. Future research should prioritize efforts to mitigate these biases and ensure accurate and precise detection in diverse populations and imaging environments.
\section{Acknowledgements}
C.I.B. is in part supported by the Helmholtz Association under the joint research school "Munich School for Data Science - MUDS".

\bibliographystyle{splncs04}
\bibliography{main}

\end{document}